| | |
|---|---|
| *Title:* | **A human visual system based 3D video quality metric** |
| *Status:* | Input Document |
| *Purpose:* | Proposal |
| *Authors:* | Amin Banitalebi-Dehkordi, Mahsa T. Pourazad, and Panos Nasiopoulos<br>2366 Main Mall,<br>Vancouver, BC, Canada<br>V6T 1Z4 |
| Tel:<br>Email: | {dehkordi, pourazad, panos}@ece.ubc.ca |
| *Source:* | University of British Columbia (UBC), TELUS Communications Inc. |

## Abstract

This contribution proposes a full-reference Human-Visual-System based 3D video quality metric. In this report, the presented metric is used to evaluate the quality of compressed stereo pair formed from a decoded view and a synthesized view. The performance of the proposed metric is verified through a series of subjective tests and compared with that of PSNR, SSIM, MS-SSIM, VIFp, and VQM metrics. The experimental results show that HV3D has the highest correlation with Mean Opinion Scores (MOS) compared to other tested metrics.

## 1 Introduction

For evaluating the quality of the compressed multi-view streams, the common Test Conditions for HEVC-based and AVC-based 3D video coding proposes to measure the PSNR of stereo pairs formed from a decoded view and a synthesized view [1][2]. PSNR and in general 2D video quality metrics do not take into account the depth effect, and the binocular properties of the human visual system (HVS). In the absence of a reliable 3D quality metric, the comparison study reported in [3] suggests the use of alternative metrics to PSNR, such as VIFp, VQM, MS-SSIM, or SSIM.

In an earlier study, we have designed a human visual-system-based quality metric for 3D videos called HV3D [4]. One of the applications of the HV3D metric is to evaluate the quality of compressed 3D videos. In this contribution, the performance of the proposed 3D video quality metric is tested on the compressed stereo pairs formed from a decoded view and a synthesized view. The results show that the HV3D metric has higher correlation with the perceived quality compared to PSNR, VIFp, VQM, MS-SSIM and SSIM.

The rest of this report is organized as follows: Section 2 provides an overview on the HV3D quality metric, Section 3 elaborates on the performance evaluation, and Section 4 concludes the report.

## 2 Human-Visual-System-based 3D (HV3D) quality metric

The proposed 3D quality metric, HV3D, takes into account the quality of individual views, the quality of the cyclopean view (fusion of the right and left view, what the viewer perceives), as well as the quality of the depth information as follows [4]:

$$HV3D = w_1 Q_{R'} + w_1 Q_{L'} + w_2 Q_{R'L'} + w_3 Q_{D'} \qquad (1)$$

where $Q_{R'}$ and $Q_{L'}$ are the quality of the distorted right and left views compared to the reference views respectively, $Q_{R'L'}$ is the quality of the cyclopean view, $Q_{D'}$ is the quality of the depth information of distorted views, and $w_1$, $w_2$, and $w_3$ are weighting constants. Figure 1 illustrates the flowchart of the proposed method.

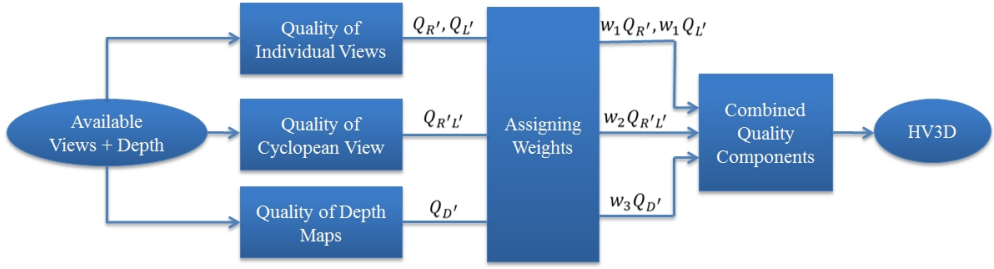

*Figure 1. Flowchart of HV3D*

## 2.1 Quality of individual views

The HV3D metric as described by equation (1) considers the quality of individual views that form the stereo pair. The quality of the distorted right view with respect to its matching reference view is calculated as follows:

$$w_1 Q_{R'} = w_1 VIF(Y_R, Y_{R'}) + w_4 VIF(U_R, U_{R'}) + w_4 VIF(V_R, V_{R'}) \qquad (2)$$

where $Y_R$ and $Y_{R'}$ are luma information of the reference and distorted right views respectively, $U_R$ and $V_R$ are the chroma information of the reference right-view, $U_{R'}$ and $V_{R'}$ are the chroma information of the distorted right-view, $w_1$ and $w_4$ are weighting constants. The quality of the left view is calculated in the same fashion.

## 2.2 Quality of cyclopean view

In order to measure the quality of the cyclopean view, first the cyclopean view is constructed by combining the corresponding areas from the left and right views. This is done by finding the matching blocks between right and left views. To this end, the luma information of each view is divided into 16×16 blocks. Note that the block size is chosen for HD video applications, to significantly reduce the complexity of the proposed approach [4]. Once the matching blocks are detected, the information of matching blocks in the left and right views needs to be fused. Here the 3D-DCT transform is applied to each pair of the matching blocks (left and right views) to generate two 16×16 DCT-blocks, which contain the DCT coefficients of the fused blocks. Since the human visual system is more sensitive to the low frequencies of the cyclopean view, only the first level of the coefficients is considered (which is a 16×16 DCT-block) and the rest of them are discarded.

Another property of the human visual system is its sensitivity to contrast. To take into account this feature, we derive a 16×16 Contrast Sensitivity Function (CSF) modeling mask and apply it to the 16×16 DCT-block so that the frequencies that are of more importance to the human visual system are assigned bigger weights. This is illustrated as follows:

$$XC = \sum_{i=1}^{16} \sum_{j=1}^{16} C_{i,j} X_{i,j} \qquad (3)$$

where $XC$ is the cyclopean-view model for a pair of matching blocks in the right and left views, $X_{i,j}$ are the low-frequency 3D-DCT coefficients of the fused view, $i$ and $j$ are the horizontal and vertical indices of coefficients, and $C_{i,j}$ is the CSF modeling mask. $C_{i,j}$ is derived based on the JPEG quantization table (see [4] for more details).

Once the cyclopean-view model for all the blocks within the distorted and reference 3D views is obtained, the quality of the cyclopean view is calculated as follows:

$$Q_{R'L'} = VIF(D, D')^\beta \sum_{i=1}^{N} \frac{SSIM(IDCT(XC_i), IDCT(XC'_i))}{N} \qquad (4)$$

where $D$ is the depth map of the reference 3D view, $D'$ is the depth map of the distorted 3D view, $XC_i$ is the cyclopean-view model for the $i^{th}$ matching block pair in the reference 3D view, $XC'_i$ is the cyclopean-view model for the $i^{th}$ matching block pair in the distorted 3D view, $IDCT$ stands for inverse 2D discrete cosine transform, $N$ is the total number of blocks in the each view, $\beta$ is a constant, empirically assigned to 0.7 (resulted from a series of subjective tests).

## 2.3 Quality of depth maps

The quality of depth map is formulated as follows:

$$Q_{D'} = VIF(D,D')^\beta \sum_{i=1}^{N} \frac{\sigma_{d_i}^2}{N.\max(\sigma_{d_i}^2 \mid i=1,2,..,N)} \tag{5}$$

where $\sigma_{d_i}$ is the variance of block $i$ in the depth map of the 3D reference view and the local disparity variance is calculated over a block size area of 64x64, since this is the area that can be fully projected onto the eye fovea when watching a 46" HD 3D display from a typical viewing distance of 3 meters [4].

## 2.4 Weighting constants

Weighting constants are found using least mean square such that the difference between the mean opinion scores (MOS) values and the HV3D values is minimized as follows:

$$\min_{w_i, i=1,2,3,4} \{\|HV3D - MOS\|^2\} \tag{6}$$

Note that the weighting constants are determined once using a training data.

## 2.5 HV3D

The maximum of HV3D occurs when the test video is identical to the reference video. To ensure that the HV3D index has maximum value equal to 1, equation (1) is divided by its maximum value. Since the maximum possible value of SSIM and VIF in equations (2), (4) and (5) is unity, $HV3D_{max}$ is equal to:

$$HV3D_{max} = 2w_1 + 4w_4 + w_2 + w_3 . \sum_{i=1}^{N} \frac{\sigma_{d_i}^2}{N.\max(\sigma_{d_j}^2 \mid j=1,2,...,N)} \tag{7}$$

The final form of the HV3D quality metric is as follows:

$$\begin{aligned}H\hat{V}3D = [&w_1 VIF(Y_R, Y_{R'}) + w_4 VIF(U_R, U_{R'}) + w_4 VIF(V_R, V_{R'}) + w_1 VIF(Y_L, Y_{L'}) + w_4 VIF(U_L, U_{L'}) + w_4 VIF(V_L, V_{L'}) \\ &+ w_2 VIF(D,D')^\beta . \sum_{i=1}^{N} \frac{SSIM(IDCT(XC_i), IDCT(XC_i'))}{N} + w_3 VIF(D,D')^\beta . \sum_{i=1}^{N} \frac{\sigma_{d_i}^2}{N.\max(\sigma_{d_j}^2 \mid j=1,2,...,N)}]/HV3D_{max}\end{aligned} \tag{8}$$

# 3 Performance Evaluation

In this report, the performance of the HV3D metric is reported for the 2-view case scenario where 2 views and their corresponding depth maps are coded and intermediate views are synthesized at the receiver side based on the decoded views and their corresponding depth maps to form stereo pairs for the multi-view display (see [5]). In our experiments the quality of the stereo pairs formed from the decoded left view and a synthesized view is evaluated with respect to the original pair (see Figure 2). Note that if the original intermediate view does not exist, this view is synthesized based on the original views and their corresponding depth maps.

To evaluate the performance of HV3D, subjective tests were performed and correlation between MOS and HV3D is calculated. Table 1 shows the test videos and synthesized views used in the experiment.

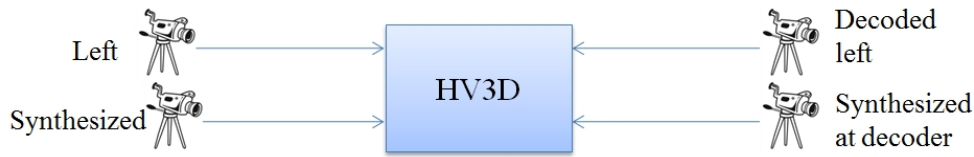

*Figure 2. Application of HV3D quality metric in the 2-view scenario*

Table 1: Input views and stereo pair for 2-view test scenario

| Seq. ID | Test Sequence | Test Class | Input views | View to synthesize | Stereo pair |
|---|---|---|---|---|---|
| S01 | Poznan_Hall2 | A | 7-6 | 6.5 | 6.5-6 |
| S02 | Poznan_Street | A | 4-3 | 3.5 | 3.5-3 |
| S03 | Undo_Dancer | A | 2-5 | 3 | 3-5 |
| S04 | GT_Fly | A | 5-2 | 4 | 4-2 |
| S05 | Kendo | C | 3-5 | 4 | 4-5 |
| S06 | Balloons | C | 3-5 | 4 | 4-5 |
| S07 | Lovebird1 | C | 6-8 | 7 | 7-8 |
| S08 | Newspaper | C | 4-6 | 5 | 5-6 |

For each test sequence, four different coding rates were studied. Specifically, the quantization parameter (QP) was set to 25, 30, 35, and 40. The test sequences were coded using the HEVC-based 3DV codec (3D-HTM 4.0). The viewing conditions for the subjective tests were set according to the ITU-R Recommendation BT.500-11 [6]. Eighteen observers participated in to the subjective tests, ranging from 21 to 29 years old. All subjects had none to marginal 3D image and video viewing experience. They were all screened for color and visual acuity (using Ishihara and Snellen charts), and also for stereo vision (Randot test – graded circle test 100 seconds of arc). The evaluation was performed using a 46" Full HD Hyundai 3D TV (Model: S465D) with passive glasses. The TV settings were as follows: brightness: 80, contrast: 80, color: 50, R: 70, G: 45, B: 30.

After a short training session, the viewers were shown the compressed and the reference stereoscopic test sequences in random order, so that the viewer would watch the reference and the compressed versions of the same sequence consecutively without knowing which video is the reference one. Between test videos, a four-second gray interval was provided to allow the viewers to rate the perceptual quality of the content and relax their eyes before watching the next video. Here, the perceptual quality reflects whether the displayed scene looks pleasant in general.

In particular, subjects were asked to rate a combination of "naturalness", "depth impression" and "comfort" as suggested by [7]. There were 10 quality levels (1-10) for ranking the videos, where score 10 indicated the highest quality and 1 indicated the lowest quality. After collecting the experimental results, outliers were removed (there were three outliers) and then the mean opinion scores from viewers were calculated.

The performance of HV3D was compared with the suggested 2D quality metrics by [3]. Figure 4 shows the logistic fitting curve associated with each of the quality metrics. Vertical axis shows the MOS and horizontal axis represents quality metrics.

Correlation of the each quality metric with MOS is calculated in terms of Pearson and Spearman Correlation Coefficients (PCC and SCC). Table 2 shows the correlation ratios for different metrics. We observe that HV3D achieves the highest correlation with the perceived quality compared to other metrics.

Table 2: Pearson and Spearman Correlation Coefficients for different quality metrics

| Metric | Spearman rank correlation coefficient (SCC) | Pearson linear correlation coefficient (PCC) |
|---|---|---|
| PSNR | 0.6357 | 0.6554 |
| SSIM | 0.6709 | 0.7034 |
| VQM | 0.6845 | 0.6805 |
| VIFp | 0.7188 | 0.7475 |
| MS-SSIM | 0.8033 | 0.7916 |
| HV3D | 0.8646 | 0.8566 |

# 4   Conclusion

In this document we present an overview of our HV3D quality metric. Performance of the proposed metric was verified through a series of subjective tests and was compared with other available quality metrics. Experimental results illustrate that the HV3D has the highest correlation with the perceived quality compared to other metrics such as PSNR, SSIM, MS-SSIM, VIFp, and VQM.

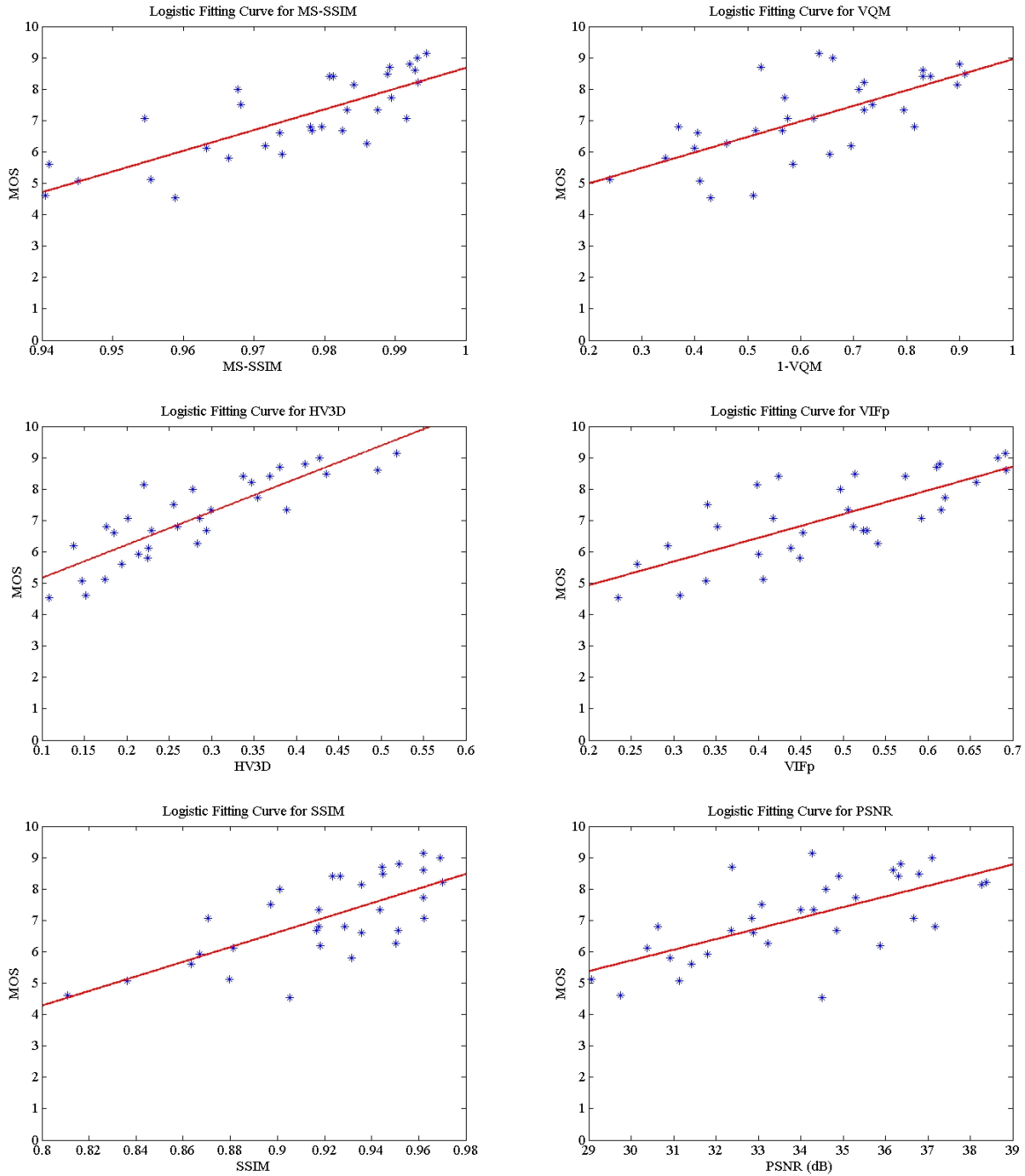

*Figure 4. Comparing the subjective results with quality metrics*

# 5 References

[1] ISO/IEC JTC1/SC29/WG11, "Common Test Conditions for HEVC- and AVC-based 3DV", Doc. N12352, Geneva, Switzerland, November 2011.

[2] ISO/IEC JTC1/SC29/WG11, "3DV: Objective quality measurement for the 2-view case scenario," Doc. M23908, San Jose, USA, February 2012.

[3] ISO/IEC JTC1/SC29/WG11, "3DV: Alternatives to PSNR," Doc. M24807, Geneva, Switzerland, May 2012.

[4] A. Banitalebi-Dehkordi, Mahsa T. Pourazad, P. Nasiopoulos, "A human visual system based 3D video quality metric," 2nd International Conference on 3D Imaging, IC3D, Dec. 2012, Liege, Belgium.

[5] ISO/IEC JTC1/SC29/WG11, "Call for Proposals on 3D Video Compression Technology", Doc. N12036, Geneva, Switzerland, March 2011.

[6] Recommendation ITU-R BT.500-11, "Methodology for the subjective assessment of the quality of the television pictures".

[7] Q. Hyunh-Thu, P. L. Callet, and M. Barkowsky, "Video quality assessment: from 2D to 3D challenges and future trends," Proc. of 2010 IEEE 17th International Conference on Image Processing, (ICIP), pp. 4025-4028, 2010.

# 6 Patent rights declaration

**Authors may have current or pending patent rights relating to the technology described in this contribution and, conditioned on reciprocity, is prepared to grant licenses under reasonable and non-discriminatory terms as necessary for implementation of the resulting ITU-T Recommendation | ISO/IEC International Standard (per box 2 of the ITU-T/ITU-R/ISO/IEC patent statement and licensing declaration form).**